# A Supervised Machine Learning Approach for Accelerating the Design of Particulate Composites: Application to Thermal Conductivity

Mohammad Saber Hashemi[1], Masoud Safdari[2], Azadeh Sheidaei[1*]

[1] Aerospace Engineering Department, Iowa State University, Ames, IA 50011, United States

[2] Aerospace Engineering Department, University of Illinois, Urbana-Champaign, IL 61820, United States

## Abstract

A supervised machine learning (ML) based computational methodology for the design of particulate multifunctional composite materials with desired thermal conductivity (TC) is presented. The design variables are physical descriptors of the material microstructure that directly link microstructure to the material's properties. A sufficiently large and uniformly sampled database was generated based on the Sobol sequence. Microstructures were realized using an efficient dense packing algorithm, and the TCs were obtained using our previously developed Fast Fourier Transform (FFT) homogenization method. Our optimized ML method is trained over the generated database and establishes the complex relationship between the structure and properties. Finally, the application of the trained ML model in the inverse design of a new class of composite materials, liquid metal (LM) elastomer, with desired TC is discussed. The results show that the surrogate model is accurate in predicting the microstructure behavior with respect to high-fidelity FFT simulations, and inverse design is robust in finding microstructure parameters according to case studies.

**Keywords:** Thermal properties, Particle-reinforced composites, Flexible composites, Computational mechanics, Machine learning

Corresponding author: Azadeh Sheidaei, PhD, Aerospace Engineering Department, Iowa State University, Ames, IA 50011, United States, Tel: 515-294-2956 (O)/Fax: 515-294-3262 (O), Email: Sheidaei@iastate.edu

## 1. Introduction

Computational material design, an emerging field of study, is a powerful technique in developing advanced multifunctional materials. Accomplishing the goal of these studies depends on the appropriate representation of the material microstructure as the design variables. Microstructure characterization and reconstruction (MCR) techniques, which are generally considered to represent the microstructure, can be categorized into (1) correlation function-based methods [1–4], (2) physical descriptor-based methods [5–8], (3) spectral density function-based characterization and reconstruction by level-cutting a random field [9] or by disk-packing [10], (4) ML-based methods such as convolutional deep neural networks [11], instance-based learning [12], and encoding/decoding methods [13], and (5) texture synthesis-based methods [14–16]. Categories 1, 4, and 5 cannot be used for material design since they do not provide specific or physical design variables. Others may involve dimensional reduction due to high-dimensional representations [17], which should be cautiously studied to avoid significant information loss and decrease the structural variability. All in all, the most convenient yet capable category for material design is the physical descriptor-based method. ML methods have been used to learn the complex relationship between microstructure descriptors and their homogenized response when dealing with a massive database. For instance, Hashemi et al. [18] recently developed a novel ML-based computational framework for homogenizing heterogeneous soft materials. Furthermore, Bessa et al. [19] have proposed a framework for data-driven analysis and material systems design under uncertainty. In such frameworks, the computational cost of high-fidelity analyses is reported as the main hurdle in the data generation phase as material analyses are inherently complex for several reasons, e.g., complexities of resolving heterogeneities of the material, non-linearity of material's response and boundary conditions, and excessive dimensionality of the design space. Reduced-order models (ROMs) could be utilized to accelerate the data generation phase. Several research works have been devoted to such developments. For example, Liu et al. [20] have developed a self-consistent clustering analysis to predict irreversible processes accurately.

In this study, we focus on designing particulate composites with LM elastomer as our case study. LM composites constitute a new class of multifunctional materials with concurrently tuned thermal, dielectric, and mechanical properties. LM composite have shown promising applications in areas such as wearable devices, electronics, robotics, and biomedical [21]. Carbon-based fibers in micron or nano-size limits the flexibility of the polymeric materials due to the huge difference in the properties with host polymer that results in a fracture within a few percent strain. However, LM elastomer stretches for 500% with 70% VF without fracture while maintaining high thermal conductivity [21]. Over the past few years, most research work has been focused on developing methods to synthesize LM droplets and their suspension within various matrix materials; for example, methods are developed for precise controlling of the size distribution and the volume fraction of LM droplets with a wide variety of surfactants, polymer coatings, and dispersion media [21–26]. As these methods are further refined, there will be an increasing need for computational tools to design LM composites with target material properties. The determination of the effective properties of composite materials given their specific constituents has been widely explored in the past decades. High-fidelity finite element (FE) simulations of composite materials' response yield accurate predictions, but the associated computational time limits their applicability in the design phase. Thus, we developed a computational framework to obtain accurate and inexpensive predictions of the TC of LM composites and understand their dependence on the microstructural geometry based on optimal ML algorithms. The overview of this material discovery framework is shown in Fig. 1. Phase 1, data generation, is necessary to train ML models over a labeled dataset. Phase 2 involves finding the complex relationship between the structure and properties using appropriate ML algorithms. The first objective of this phase is finding the effective properties of a material system given its microstructural parameters, such as volume fraction (VF), size distribution, and aspect ratio (AR). The second objective is the inverse design, i.e., finding microstructure parameters given its properties as inputs. The microstructure could be realized, given its parameters or 3-D visualization through computational packing algorithms in the studied material system. Thus, Phase 3 encompasses the inference of direct structure-property relationships and generating and visualizing candidate microstructures from the inverse design framework. The computational times reported in this study are based on what has been achieved on a machine with AMD Ryzen 1950x CPU and 32 GB DDR4 RAM.

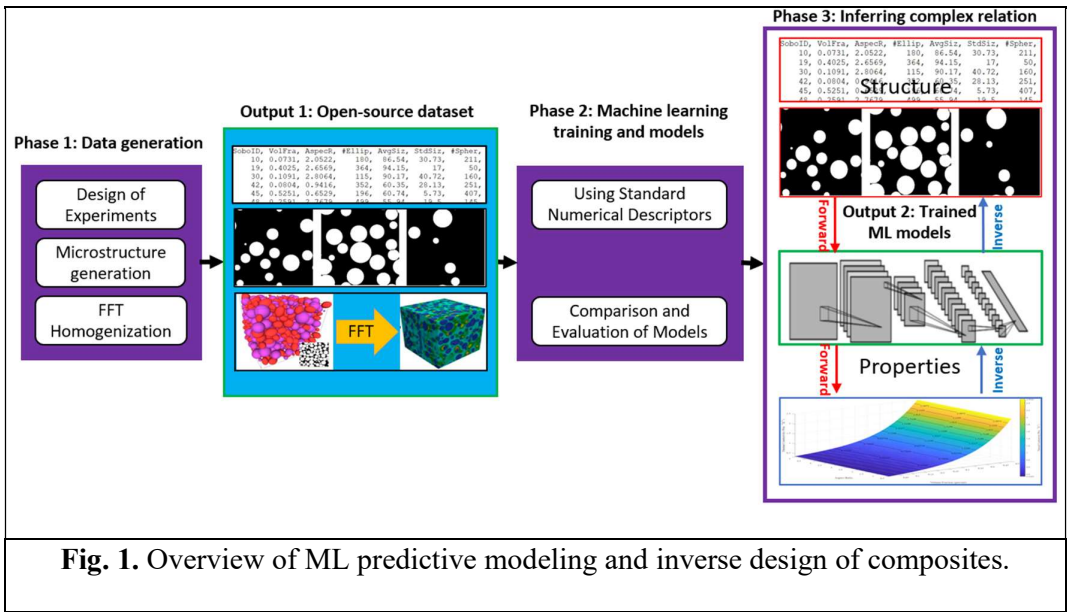

**Fig. 1.** Overview of ML predictive modeling and inverse design of composites.

## 2. Methods

### 2.1. Data generation

To have a sufficiently large dataset for advanced ML algorithms, we cannot only rely on the experimental results of the material. Even best designed experimental procedures cannot cover all feature vectors, material system characteristics in this study, required for the ML training as diverse and representative as their computational counterparts. Therefore, a robust and efficient design scheme of virtual experiments, i.e., computational simulations, was necessary to have a representative dataset of the random variables, which were the physical microstructure descriptors in this study, affecting the performance of a trained ML predictive model. The output of this phase is an open-source dataset that could open new horizons for research on the material discovery, mainly when it can be applied to similar material systems of particulate composites with different material constituents.

### 2.1.1. Design of experiment

The target variables to be predicted are an effective TC tensor of a composite given a specified set of materials for the composite constituents. The properties will be computationally measured using a homogenization technique. Since the effective TC of microstructure can be calculated based on the linear equation of heat conduction, assuming constant material coefficients, a set of constant periodic boundary conditions can be prescribed to perform homogenization. Thus, the only remaining parameter that affects the property is the microstructural morphology. Since the studied composite was insulated LM elastomer, MCR type we chose was based on the physical descriptors. It has been shown [21–26] that LM particulates are roughly encapsulated in an ellipsoidal shape with varying aspect ratios ($AR$s), following a normal distribution. To account for different $AR$s, two $AR$s for particles were considered: i) $AR = 1$ for spherical particles, and ii) $AR \neq 1$ for all other ellipsoidal particles. Another important geometrical factor in composites is the volume fraction ($VF$) of the constituents. The physical descriptors and their bounds, as well as the numbers of particles, which are necessary for packing algorithm performance, are summarized in Table 1 where $VF(\%)$ denotes the volume fraction, $AR$ denotes the ellipsoidal aspect ratio, $Avg(\mu m)$ is the average particle size, $Std(\mu m)$ is the standard deviation of the particle sizes, $\#Ell$ is the number of ellipsoidal particles, and $\#Sph$ is the number of spherical particles. The bounds of variables were selected based on an experimental work [24].

**Table 1.** The bounds on the physical descriptors of the microstructure and the numbers of particles inside a pack.

|  | $VF(\%)$ | $AR$ | $\#Ell$ | $Avg(\mu m)$ | $Std(\mu m)$ | $\#Sph$ |
|---|---|---|---|---|---|---|
| Lower Bounds | 1 | 0.5 | 1 | 0.1 | 0.1 | 1 |
| Upper Bounds | 60 | 3.0 | 500 | 100.0 | 100.0 | 500 |

After identifying and limiting the microstructure parameters affecting the properties, a DoE method was used to explore the design or input variables' domain for training machine learners efficiently. Since there was no prior knowledge of the conditional probabilities of the microstructural inputs and the property output, space-filling designs that equally cover different regions of the design space were chosen. Different optimum Latin Hypercube Samplings [27] and the Sobol sequence [28], a deterministic low discrepancy sequence, have shown a better balance between more regular distribution or randomness and being closer to a regular grid or better coverage of the input variables space [29]. Thus, we chose the Sobol sequence, which is also very fast in generating the experiment points.

### 2.1.2. Microstructure generation

Rocpack [30], a derivative of the Lubachevsky-Stillinger (LS) algorithm [31] for packing disks and spheres, was used to generate a microstructure for each point of DoE. It allows the specification of material characterization parameters presented in Section 2.1.1 and different realizations of a microstructure with unique parameters by a random initial seeding of particles. Not all experiment points generated by the Sobol sequence are consistent or physically meaningful. For instance, the continuous normal size distribution of the particles would be discretized according to the total number of particles resulting in different growth rates and final sizes for the particles. The minimum and maximum sizes of particles can be consequently determined by

$$
\begin{aligned}
&\text{If } Expr \geq 0 \Longrightarrow \begin{cases} r_{min} = Avg - \sqrt{Expr} \\ r_{max} = Avg + \sqrt{Expr} \end{cases} PDF(r) = \frac{1}{Std\sqrt{2\pi}} \exp\left(-\frac{1}{2}\left(\frac{r-Avg}{Std}\right)^2\right) \\
&\xRightarrow{PDF=\frac{1}{N}} (r - Avg)^2 = -2Std^2 \ln\left(\frac{Std}{N}\sqrt{2\pi}\right) = Expr
\end{aligned}
\tag{1}
$$

where $PDF$ and $N$ are the Gaussian probability distribution function and the number of particles, respectively. This equation is obtained through rearranging the distribution function, and if the minimum size $r_{min}$ is lower than zero or $Expr$ is lower than zero, the parameters are physically inconsistent. The numbers of particles also determine the dimension of the microstructure in a periodic cube. This can be inferred from Eq. (2), which elucidates the implicit relationship between the physical descriptors of the microstructure by expanding the volume fraction in terms of the size distribution (PDF).

$$
\begin{aligned}
&Domain_{Size} = Vol_{Total}^{\frac{1}{3}} = \left(\frac{Vol_{Inclusions}}{VF}\right)^{\frac{1}{3}} = \left(\frac{Vol_{Ellipsoids}+Vol_{Spheres}}{VF}\right)^{\frac{1}{3}} = \\
&\left\{\frac{1}{VF}\left(\int_{R_{min}}^{R_{max}}[4\pi AR r^2 N_{Ellipsoids} PDF(r)]dr + \right.\right. \\
&\left.\left. \int_{R_{min}}^{R_{max}}[4\pi r^2 N_{Spheres} PDF(r)]dr\right)\right\}^{\frac{1}{3}}
\end{aligned}
\tag{2}
$$

The outputs of the code, i.e., 3D realization of the packs, were given as 2-D images of the sliced 3-D microstructures in one arbitrary direction due to the isotropy. Based on the uniform distribution for particle orientations in packing and our high-fidelity FFT analyses, off-diagonal thermal conductivity values were orders of magnitude smaller than the diagonal values, and the diagonal values were also close to mean value for thermal conductivity of the tensor which further assures our samples were close to isotropic. The resolution of slicing limits voxelization, and it cannot be arbitrarily increased since the FFT homogenization cost scales super-linearly with the number of voxels used. Therefore, we set the number of pixels in all directions to 300. This setting may be coarse for packs with tiny LM particles to capture the exact geometrical shape of the particles, but it resulted in an average FFT computation time of 3 hours for each pack. We also generated a few packs with higher resolutions. The results of homogenized property did not change after 300 pixels significantly. Therefore, we chose the minimum required resolution to minimize the computational cost of data generation in FFT homogenization step.

### 2.1.3.    FFT homogenization to calculate effective TC

TC values of 0.29W/mK and 26.4W/mK were selected for the silicone elastomer matrix and eutectic gallium-indium (EGaIn), respectively. Conventional numerical methods such as Finite element for finding the effective properties of random heterogeneous materials suffer from their dependency on very fine and high-quality mesh conforming to intricate geometries of phases. The FFT method is shown to be efficient with voxelized representative volume elements (RVE) as no conformal meshing is required [32]. It is also superior to other numerical methods in terms of scalability $O(NlogN)$ in complexity vs. $O(N^3)$. In a separate study, we have validated this homogenization method with the experimental results of the LM composites [33], and the reader is referred to this work for more details.

## 2.2. *ML model training*

Phase 2 aims to find an efficient and optimal ML model to replace the time-consuming homogenization process. Neural networks are versatile and robust as a regression tool for modeling complex functions. Each neuron can be a nonlinear function and using different network architectures, number of neurons, number of layers, and links between neurons may arbitrarily increase their complexity. Therefore, we considered different architectures and trained them on the dataset according to the n-fold cross-validation technique. Although ReLU units generally perform better when using data ranging outside the regular interval [-1, 1], the perceptron function was Sigmoid. Therefore, the input data have been linearly normalized into [-1, 1]. The inputs were vectors of physical descriptors and other packing parameters needed for microstructure reconstruction. Simultaneously, the only output was the homogenized TC, which was the average value of the diagonal elements of the TC tensor, assuming the material system under study is almost isotropic. The whole available dataset of homogenized packs was randomly divided into five equally sized sections. The neural networks were trained five times by using a section of data, which has not been considered previously, as a test set and the rest as a training set each time. After five training processes, the average training accuracy and its standard deviation were calculated so that the performance of different architectures could be compared. The best performing network with the highest average training accuracy was chosen for the final training on the whole dataset.

*2.3. Inferring complex structure-property relationship*

Based on Section 2.2, a fast and reliable ML model for material properties prediction can be found to act as a surrogate of relatively expensive direct numerical solvers and to establish the direct relationship between the structure of the studied particulate material and its effective homogenized properties. However, the more demanding problem is the inverse design, which has been challenging due to inefficient and expensive methods of finding the material properties for a given microstructure, especially when dealing with the complex characterization of microstructure images with too many features. Since our studied material system could be characterized by only six features (or physical descriptors), and we have already established a reliable yet fast surrogate model for the direct structure-property relationship discussed in Section 2.2, an elitist genetic algorithm (GA) was utilized to optimize the structure according to limits imposed in experimental studies, e.g., the lower and upper bounds mentioned in Table 1 based on [24]. The algorithm begins with a random initial population, which consists of several candidate points in the design space. It will then continue generating new populations based on the previous ones iteratively until one prespecified stopping criteria is met. After each population generation, the objective function is evaluated for each member of population to determine the member fitness, a scale of being more optimum or having higher survival rate in the next population. Based on their fitness, some members will be passed as elites to the next generation if they have the best fitness or minimum objective values; otherwise, new members will be created through genetic operators of mutation (random change in the vector of single parent member) and crossover (combining the vectors of two parent members). The main prohibiting factor in evolutionary algorithms is the computational complexity due to the fitness calculation of many design points in each iteration [34]; however, the objective function in our study was calculated by the trained neural network, which is very fast in inference. The single target of such an optimization is the isotropic TC of the composite, and the design variables are the physical descriptors of Section 2.1. The objective function was selected to be the absolute difference between the ML prediction and the desired property to cast the problem into a minimization form. The flowchart of the optimization method of inverse material design is shown in Fig. 2. A green color distinguishes the evaluation step with the trained surrogate model.

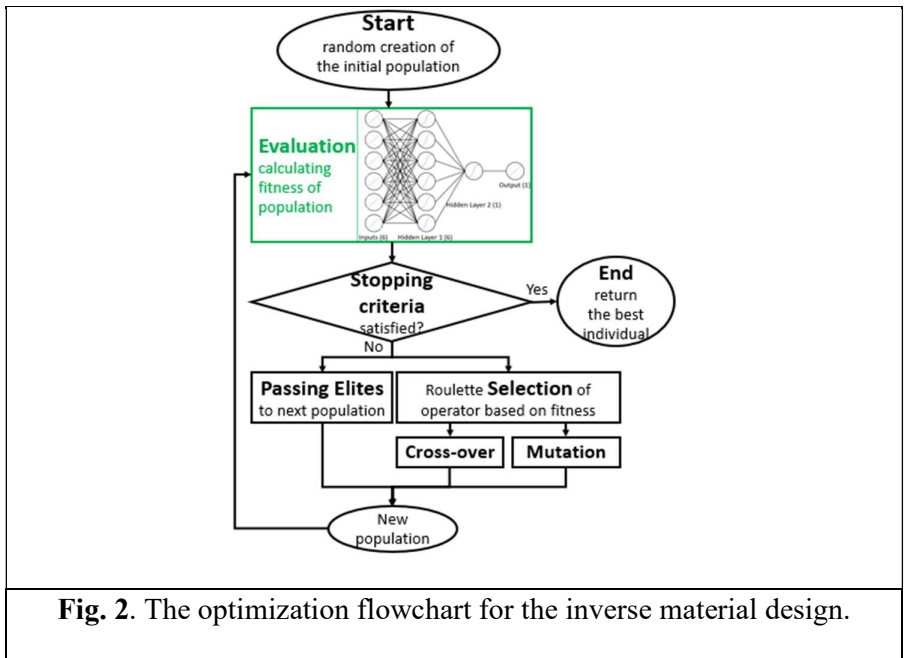

**Fig. 2**. The optimization flowchart for the inverse material design.

## 3. Results and Discussion

### 3.1. Generated database

A subset of the first DoE results of almost 10,000 points based on the Sobol sequence is given in Table 2. In this table, SID denotes the Sobol ID or the position of the parameters in the sequence, and columns 2-7 are needed for microstructure generation. The last columns contain minimum and maximum radii of the ellipsoid and spherical particles, and the domain size of the microstructure, which is required for the packing algorithm, respectively. The unit of all dimensions is micrometer ($\mu m$). Naturally, Sobol IDs should be 1, 2, 3, …, and the absent Sobol IDs are due to physically inconsistent sets of parameters or others for which packs were not generated under an hour time limit as discussed in Section 2.1.2.

**Table 2.** The first Sobol-based inputs of the material structure database; columns 2-7 are the features of the neural network, while the last column can be calculated from them by Eq. (2).

| SID | VF | AR | #Ell | Avg | Std | #Sph | $R_{min}^{El}$ | $R_{max}^{El}$ | $R_{min}^{Sph}$ | $R_{max}^{Sph}$ | DS |
|-----|------|-----|------|------|------|------|---------|---------|----------|----------|-------|
| 10 | 7.3 | 2.1 | 180 | 86.5 | 30.7 | 211 | 46.5 | 126.6 | 42.9 | 130.2 | 866.2 |
| 19 | 40.3 | 2.6 | 364 | 94.2 | 17.0 | 50 | 58.9 | 129.4 | 84.5 | 103.8 | 646.4 |
| 30 | 10.9 | 2.8 | 115 | 90.2 | 40.7 | 160 | 70.3 | 110.1 | 51.6 | 128.8 | 602.6 |

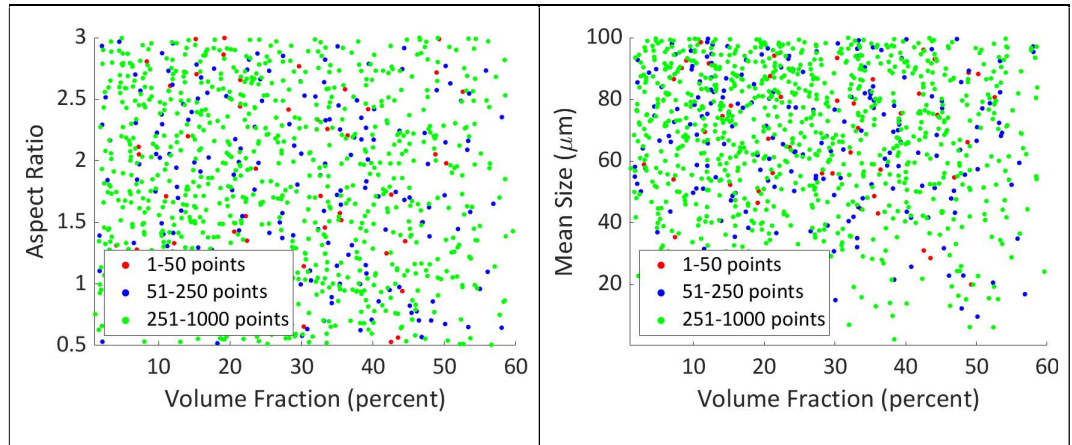

**Fig. 3.** The first 1000 feasible DoE points generated by the Sobol sequence are projected on different 2D planes.

Space-filling designs should cover the design space almost homogeneously while they need to maintain non-collapsing constraints. From Fig. 3 of the generated packs, it can be inferred that the criteria are met for our Sobol DoE although some generated sets of parameters were not used in the final simulations due to physical inconsistencies or some long times needed for packing. This design has an advantage of successive coverage of space along with the sequence generation so that the dataset can be successively improved, i.e., the design space can be further explored by continuing the previous number sequences. For instance, the first 50 DoE points of generated packs were shown by a red color, then the next 200 and the next 750 points were plotted by blue and green colors, respectively. Additionally, the projection of 6D points on different 2D planes, VF-Mean Size and VF-AR, did not overlap each other.

The 3D visualization of a microstructure pack (purple spherical particles and yellow ellipsoidal ones) and a sample of its 2D slices (a black-and-white image) are shown in the top part of Fig. 4. The respective FFT simulation results (thermal gradient) due to thermal loading of ΔT=[0,10,0] are shown at the bottom of Fig. 4. The gradient is larger in high-concentration regions.

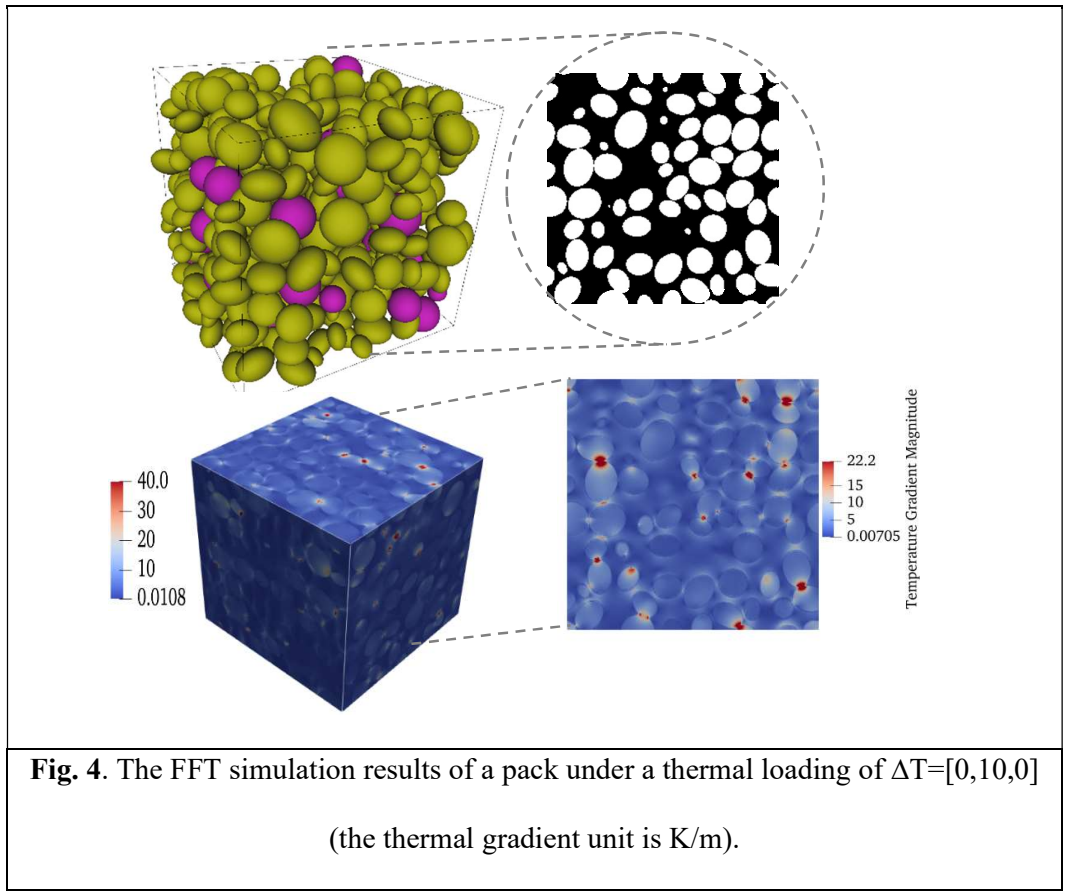

**Fig. 4**. The FFT simulation results of a pack under a thermal loading of ΔT=[0,10,0] (the thermal gradient unit is K/m).

*3.2. Optimized surrogate model of direct structure-property relationship*

As stated in Section 2, several fully connected neural network architectures were considered, trained, and compared to find a network with high expected prediction accuracy. Their characteristics are described in Table 3.

**Table 3**. Neural network architectures grid-searched for ML hyperparameter

optimization.

| Network type | Layer | Number of neurons | | | | | |
|---|---|---|---|---|---|---|---|
| 1-layer | Layer 1 | 3 | 6 | 10 | 20 | 50 | 100 |
| 2-layer | Layer 1 | 3 | 6 | 10 | 20 | 50 | 100 |
| | Layer 2 | 1 | 6 | 10 | 20 | 50 | 100 |

According to the cross-validation technique, the best network with the lowest mean squared error (MSE) is shown in Fig. 5(a). Its regression plot over the whole dataset, Fig. 5(b), shows that the accuracy is lower for large conductivity composites due to fewer DoE points covering regions of design space with higher VFs. Furthermore, the error histogram in the subplot of Fig. 5 (b) indicates that most errors with respect to the homogenized packs are relatively small. It is worth mentioning that the speed of the surrogate model in terms of a trained neural network is an order of 0.1 seconds compared with the conventional method of packing plus homogenization, which took an average time of 1+3 hours for each microstructure in our developed database.

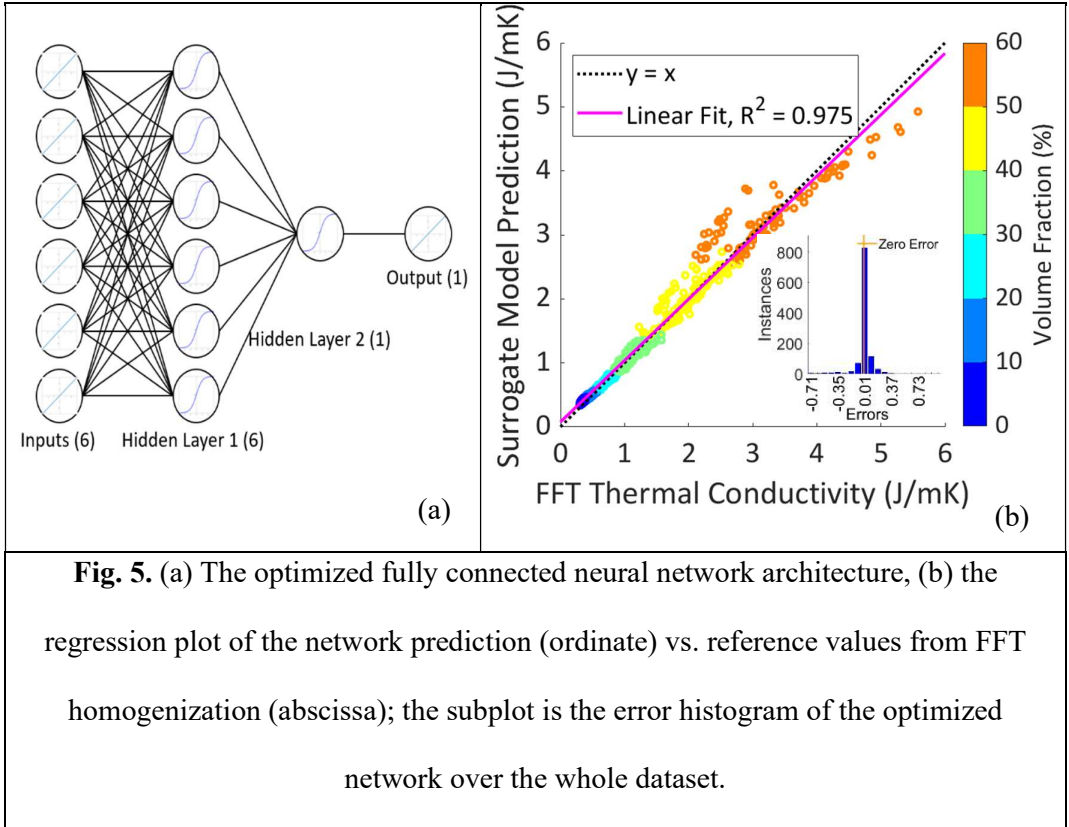

**Fig. 5.** (a) The optimized fully connected neural network architecture, (b) the regression plot of the network prediction (ordinate) vs. reference values from FFT homogenization (abscissa); the subplot is the error histogram of the optimized network over the whole dataset.

Following our objective of inferring the direct relationship between microstructure and its properties, several response surfaces of the studied LM composite were plotted using the trained neural network. For each surface, all microstructure features, network inputs, were fixed except for two of them, warm colors show high conductivity composites, and black lines on the surface are constant TC contours. Fig. 6(a) is the response surface of TC-AR-VF.

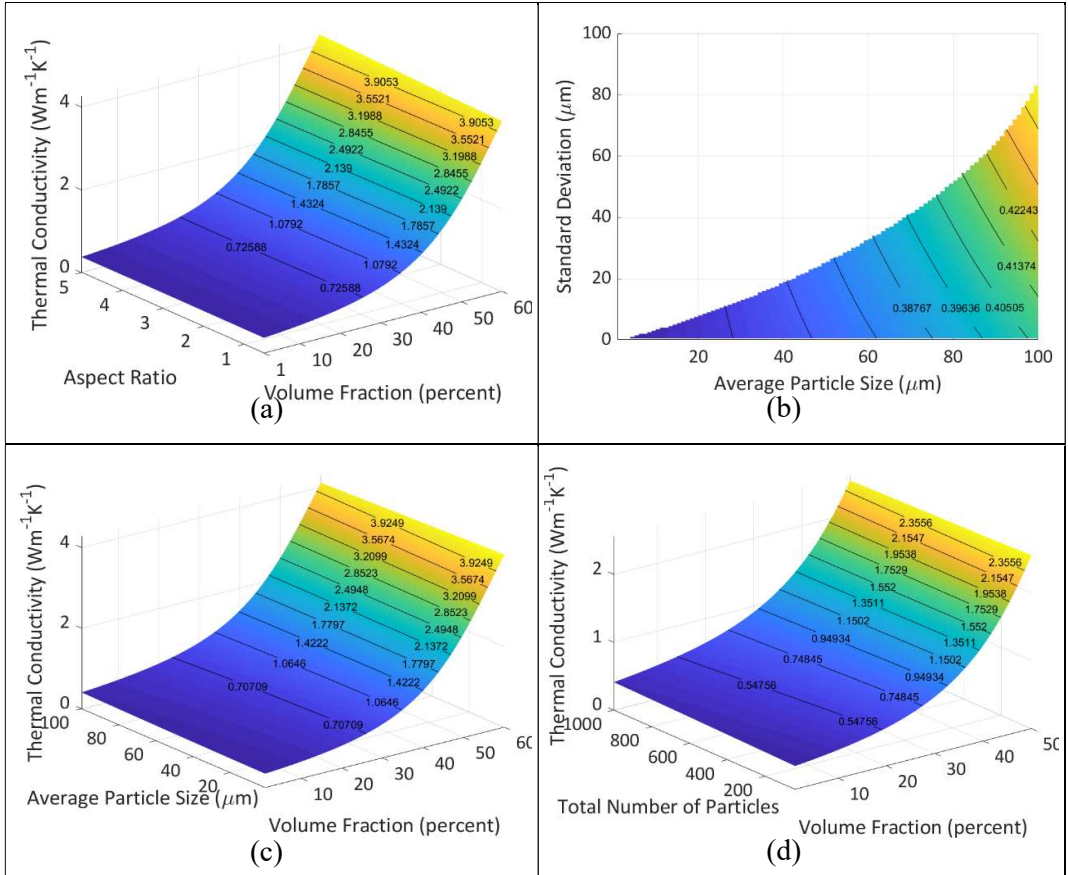

**Fig. 6.** Inferring the direct Structure-Property relationship of the studied LM composite via plotting the response surfaces: (a) TC-VF-AR, (b) TC-Avg-Std projection on Avg-Std plane, (c) TC-VF-Avg, and (d) TC-VF-Number of Particles.

As expected, VF has a prominent effect on the property although very high ARs, which were not studied in this work based on experimental results, may play an important role in determining material property. Since some design points are not feasible, as pointed out in Eq. (1), the projection of the TC-Avg-Std surface on the Avg-Std plane is empty in some regions of Fig. 6(b). Thus, it is deduced that the TC has a direct relationship with the Avg and Std in particle sizes; however, their effects are much less significant than that of VF. The effect of VF and Avg is illustrated in Fig. 6(c). Again, VF is shown to be the most critical factor in TC, and Avg has a negligible effect on TC. Fig. 6(d) shows that TC has been almost constant with an increasing number of particles. It is satisfactory in this study in that the standard deviation of predicted property due to the variation in microstructure size was low. In other words, the calculated domain sizes based on Eq. (2) were sufficient to define RVE sizes.

### 3.3. Inverse design via GA optimization

A case study was done to show how the proposed method in Section 2.3 performs and to verify its results through the data generation process discussed in Section 2.1. The goal of optimization was set to 3 W/mK heat conductivity. The best design point among the last population as well as the predicted property value and the FFT calculated one are [0.522, 1.61, 223, 61.04, 69.37, 238], 2.9, and 2.7, respectively. The error may be due to the limited resolution of reconstruction and smaller surrogate model prediction errors. Following multiple tests, it can be concluded that the inverse design method is efficient and accurate enough. Additionally, the total inverse design optimization took 1 min on average. To emphasize the efficiency of our proposed computational framework, a summary of computational times is presented in Table 4. If there was no surrogate model, each design point in the inverse design optimization loop would have taken around 4 hours (packing+FFT homogenization) to be objectively quantified, while it is taking seconds using our trained surrogate model.

Table 4. Average computational time for different processes in minutes.

| Packing algorithm/RVE | FFT homogenization/RVE | Surrogate model training and prediction/RVE | Inverse design |
|---|---|---|---|
| 60 | 180 | 1 | 2 |

## 4. Conclusions

This paper proposed a new supervised machine learning approach for accelerating the prediction of the thermal conductivity of the particulate composites, as well as designing a composite with the desired property. This framework has the advantages of superior computational speed compared to conventional optimization techniques. A comprehensive database for particulate composites has been generated covering the whole design space. The microstructure reconstructions based on this study's DoE can also be used for similar heterogeneous particulate materials with different constituents. Additionally, a surrogate ML model was trained on the database to establish the direct links between the microstructure and the conductivity property and visualize them with various response surfaces in minutes compared with days for the traditional method of microstructure reconstruction and direct numerical solution. For the studied material system, the VF is far more important in determining the conductivity; however, greater particle sizes and higher ARs slightly improve TC. The smart and physically aware choice of the specified physical descriptors for MCR not only provided less-complicated modeling of structure-property links with respect to the image-based convolution neural networks which require many more training data but also connected the results of this study directly to the process phase, which is readily prepared for material synthetization. Finally, the low number of characterization features, the target TC, and a fully connected neural network as the fast surrogate model trained on our generated database enabled us to use an evolutionary optimization, GA, to explore the design space and find the physical descriptors of an LM composite which will have a given TC in about a minute.

**Data availability**: Supplementary data to this article can be found online at https://github.com/ms-hashemi/Insulated-LM-elastomer-conductivity

**Declaration of Competing Interest**: The authors declare that there are no conflicts of interest.

**Acknowledgement:** This research has been funded by Iowa State University.

## References

[1]     A. Sheidaei, M. Baniassadi, M. Banu, P. Askeland, M. Pahlavanpour, N. Kuuttila, F. Pourboghrat, L.T. Drzal, H. Garmestani, 3-D microstructure reconstruction of polymer nano-composite using FIB-SEM and statistical correlation function, Composites Science and Technology. 80 (2013) 47–54. https://doi.org/10.1016/j.compscitech.2013.03.001.

[2]     S.A. Tabei, A. Sheidaei, M. Baniassadi, F. Pourboghrat, H. Garmestani, Microstructure reconstruction and homogenization of porous Ni-YSZ composites for temperature dependent properties, Journal of Power Sources. 235 (2013) 74–80. https://doi.org/10.1016/j.jpowsour.2013.02.003.

[3]     H. Amani Hamedani, M. Baniassadi, A. Sheidaei, F. Pourboghrat, Y. Rémond, M. Khaleel, H. Garmestani, Three-Dimensional Reconstruction and Microstructure Modeling of Porosity-Graded Cathode Using Focused Ion Beam and Homogenization Techniques, Fuel Cells. 14 (2014) 91–95. https://doi.org/10.1002/fuce.201300170.

[4]     C.L.Y. Yeong, S. Torquato, Reconstructing random media, Physical Review E - Statistical Physics, Plasmas, Fluids, and Related Interdisciplinary Topics. 57 (1998) 495–506. https://doi.org/10.1103/PhysRevE.57.495.

[5]     H. Xu, Y. Li, C. Brinson, W. Chen, A descriptor-based design methodology for developing heterogeneous microstructural materials system, Journal of Mechanical Design, Transactions of the ASME. 136 (2014). https://doi.org/10.1115/1.4026649.

[6]     G. Requena, G. Fiedler, B. Seiser, P. Degischer, M. di Michiel, T. Buslaps, 3D-Quantification of the distribution of continuous fibres in unidirectionally reinforced composites, Composites Part A: Applied Science and Manufacturing. 40 (2009) 152–163. https://doi.org/10.1016/j.compositesa.2008.10.014.

[7]     H. You, Y. Kim, G.J. Yun, Computationally fast morphological descriptor-based microstructure reconstruction algorithms for particulate composites, Composites Science and            Technology.            182            (2019)            107746. https://doi.org/10.1016/j.compscitech.2019.107746.

[8]     E. Yousefi, A. Sheidaei, M. Mahdavi, M. Baniassadi, M. Baghani, G. Faraji, Effect of nanofiller geometry on the energy absorption capability of coiled carbon nanotube composite material, Composites Science and Technology. 153 (2017) 222–231. https://doi.org/10.1016/j.compscitech.2017.10.025.

[9]     J.W. Cahn, Phase separation by spinodal decomposition in isotropic systems, The Journal of Chemical Physics. 42 (1965) 93–99. https://doi.org/10.1063/1.1695731.

[10]     S. Yu, Y. Zhang, C. Wang, W.K. Lee, B. Dong, T.W. Odom, C. Sun, W. Chen, Characterization and Design of Functional Quasi-Random Nanostructured Materials Using Spectral Density Function, Journal of Mechanical Design, Transactions of the ASME. 139 (2017). https://doi.org/10.1115/1.4036582.

[11]     R. Cang, Y. Xu, S. Chen, Y. Liu, Y. Jiao, M.Y. Ren, Microstructure Representation and Reconstruction of Heterogeneous Materials Via Deep Belief Network for Computational Material Design, Journal of Mechanical Design, Transactions of the ASME. 139 (2017). https://doi.org/10.1115/1.4036649.

[12]    V. Sundararaghavan, N. Zabaras, Classification and reconstruction of three-dimensional microstructures using support vector machines, Computational Materials Science. 32 (2005) 223–239. https://doi.org/10.1016/j.commatsci.2004.07.004.

[13]    R. Bostanabad, A.T. Bui, W. Xie, D.W. Apley, W. Chen, Stochastic microstructure characterization and reconstruction via supervised learning, Acta Materialia. 103 (2016) 89–102. https://doi.org/10.1016/j.actamat.2015.09.044.

[14]    L.Y. Wei, M. Levoy, Fast texture synthesis using tree-structured vector quantization, in: Proceedings of the ACM SIGGRAPH Conference on Computer Graphics, Association for Computing Machinery (ACM), New York, New York, USA, 2000: pp. 479–488. https://doi.org/10.1145/344779.345009.

[15]    V. Sundararaghavan, Reconstruction of three-dimensional anisotropic microstructures from two-dimensional micrographs imaged on orthogonal planes, Integrating Materials and Manufacturing Innovation. 3 (2014) 240–250. https://doi.org/10.1186/s40192-014-0019-3.

[16]    R. Bostanabad, Y. Zhang, X. Li, T. Kearney, L.C. Brinson, D.W. Apley, W.K. Liu, W. Chen, Computational microstructure characterization and reconstruction: Review of the state-of-the-art techniques, Progress in Materials Science. 95 (2018) 1–41. https://doi.org/10.1016/j.pmatsci.2018.01.005.

[17]    Z. Yang, X. Li, L.C. Brinson, A.N. Choudhary, W. Chen, A. Agrawal, Microstructural materials design via deep adversarial learning methodology, Journal of Mechanical Design, Transactions of the ASME. 140 (2018). https://doi.org/10.1115/1.4041371.

[18]    M.S. Hashemi, M. Baniassadi, M. Baghani, D. George, Y. Remond, A. Sheidaei, A novel machine learning based computational framework for homogenization of

heterogeneous soft materials: application to liver tissue, Biomechanics and Modeling in Mechanobiology. 19 (2020) 1131–1142. https://doi.org/10.1007/s10237-019-01274-7.

[19]    M.A. Bessa, R. Bostanabad, Z. Liu, A. Hu, D.W. Apley, C. Brinson, W. Chen, W.K. Liu, A framework for data-driven analysis of materials under uncertainty: Countering the curse of dimensionality, Computer Methods in Applied Mechanics and Engineering. 320 (2017) 633–667. https://doi.org/10.1016/j.cma.2017.03.037.

[20]    Z. Liu, M.A. Bessa, W.K. Liu, Self-consistent clustering analysis: An efficient multi-scale scheme for inelastic heterogeneous materials, Computer Methods in Applied Mechanics and Engineering. 306 (2016) 319–341. https://doi.org/10.1016/j.cma.2016.04.004.

[21]    M.D. Bartlett, N. Kazem, M.J. Powell-Palm, X. Huang, W. Sun, J.A. Malen, C. Majidi, High thermal conductivity in soft elastomers with elongated liquid metal inclusions, Proceedings of the National Academy of Sciences of the United States of America. 114 (2017) 2143–2148. https://doi.org/10.1073/pnas.1616377114.

[22]    R. Tutika, S.H. Zhou, R.E. Napolitano, M.D. Bartlett, Mechanical and Functional Tradeoffs in Multiphase Liquid Metal, Solid Particle Soft Composites, Advanced Functional Materials. 28 (2018) 1804336. https://doi.org/10.1002/adfm.201804336.

[23]    C. Pan, E.J. Markvicka, M.H. Malakooti, J. Yan, L. Hu, K. Matyjaszewski, C. Majidi, A Liquid-Metal–Elastomer Nanocomposite for Stretchable Dielectric Materials, Advanced Materials. 31 (2019) 1900663. https://doi.org/10.1002/adma.201900663.

[24]    R. Tutika, S. Kmiec, A.B.M. Tahidul Haque, S.W. Martin, M.D. Bartlett, Liquid Metal-Elastomer Soft Composites with Independently Controllable and Highly Tunable Droplet Size and Volume Loading, ACS Applied Materials and Interfaces. 11 (2019) 17873–17883. https://doi.org/10.1021/acsami.9b04569.

[25]     M.H. Malakooti, N. Kazem, J. Yan, C. Pan, E.J. Markvicka, K. Matyjaszewski, C. Majidi, Liquid Metal Supercooling for Low-Temperature Thermoelectric Wearables, Advanced Functional Materials. 29 (2019) 1906098. https://doi.org/10.1002/adfm.201906098.

[26]     H. Wang, Y. Yao, Z. He, W. Rao, L. Hu, S. Chen, J. Lin, J. Gao, P. Zhang, X. Sun, X. Wang, Y. Cui, Q. Wang, S. Dong, G. Chen, J. Liu, A Highly Stretchable Liquid Metal Polymer as Reversible Transitional Insulator and Conductor, Advanced Materials. 31 (2019) 1901337. https://doi.org/10.1002/adma.201901337.

[27]     M.E. Johnson, L.M. Moore, D. Ylvisaker, Minimax and maximin distance designs, Journal of Statistical Planning and Inference. 26 (1990) 131–148. https://doi.org/10.1016/0378-3758(90)90122-B.

[28]     I.M. Sobol, Uniformly distributed sequences with an additional uniform property, USSR Computational Mathematics and Mathematical Physics. 16 (1976) 236–242. https://doi.org/10.1016/0041-5553(76)90154-3.

[29]     J. Santiago, M. Claeys-Bruno, M. Sergent, Construction of space-filling designs using WSP algorithm for high dimensional spaces, Chemometrics and Intelligent Laboratory Systems. 113 (2012) 26–31. https://doi.org/10.1016/j.chemolab.2011.06.003.

[30]     G. Amadio, T.L. Jackson, A New Packing Code for Creating Mirostructures of Propellants and Explosives, in: 51st AIAA/SAE/ASEE Joint Propulsion Conference, American Institute of Aeronautics and Astronautics, Reston, Virginia, 2015. https://doi.org/10.2514/6.2015-4098.

[31]     B.D. Lubachevsky, F.H. Stillinger, Geometric properties of random disk packings, Journal of Statistical Physics. 60 (1990) 561–583. https://doi.org/10.1007/BF01025983.

[32]     H. Moulinec, P. Suquet, A numerical method for computing the overall response of nonlinear composites with complex microstructure, Computer Methods in Applied Mechanics and Engineering. 157 (1998) 69–94. https://doi.org/10.1016/S0045-7825(97)00218-1.

[33]     F. Nosouhi Dehnavi, M. Safdari, K. Abrinia, A. Sheidaei, M. Baniassadi, Numerical study of the conductive liquid metal elastomeric composites, Materials Today Communications. 23 (2020) 100878. https://doi.org/10.1016/j.mtcomm.2019.100878.

[34]     J. Cohoon, J. Kairo, J. Lienig, Evolutionary Algorithms for the Physical Design of VLSI Circuits, in: Springer, Berlin, Heidelberg, 2003: pp. 683–711. https://doi.org/10.1007/978-3-642-18965-4_27.